\documentstyle[12pt,epsfig]{article}
\newcommand{\bara}{\begin{array}{c}}
\newcommand{\eara}{\end{array}}

\def\be{\begin{equation}}
\def\ee{\end{equation}}
\def\ba{\begin{eqnarray}}
\def\ea{\end{eqnarray}}
\def\bann{\begin{eqnarray*}}
\def\eann{\end{eqnarray*}}
\def\benn{\begin{displaymath}}
\def\eenn{\end{displaymath}}
\def\nn{\nonumber}
\def\lapproxeq{{\ \lower 0.6ex \hbox{$\buildrel<\over\sim$}\ }}
\def\gapproxeq{{\ \lower 0.6ex \hbox{$\buildrel>\over\sim$}\ }}
\begin{document}
\vspace*{-2cm}
\renewcommand{\thefootnote}{\fnsymbol{footnote}}
\begin{flushright}
hep-ph/9909271\\
DTP/99/92\\
September 1999\\
\end{flushright}
\vskip 65pt
\begin{center}
{\Large\bf   On the Resummation of Subleading Logarithms
 in the Transverse Momentum Distribution
 of  Vector Bosons
 Produced at Hadron Colliders}\\
\vspace{1.2cm}
{
Anna~Kulesza${}^1$\footnote{Anna.Kulesza@durham.ac.uk} and
W.~James~Stirling${}^{1,2}$\footnote{W.J.Stirling@durham.ac.uk}
}\\
\vspace{10pt}
{\sf 1) Department of Physics, University of Durham,
Durham DH1 3LE, U.K.\\

2) Department of Mathematical Sciences, University of Durham,
Durham DH1 3LE, U.K.}

\vspace{70pt}
\begin{abstract}
The perturbation series for electroweak vector boson production at {\mbox small}
transverse momentum is dominated by large double logarithms at each
order in perturbation theory. An accurate theoretical prediction
therefore requires a resummation of these logarithms.
This can be performed either directly in transverse momentum space
or in impact parameter (Fourier transform) space. While both approaches
resum the same {\it leading} double logarithms, the {\it subleading} logarithms
are, in general, treated differently. We comment on two recent approaches
to this problem, emphasising the particular subleading logarithms
resummed in each case and the numerical differences in the cross sections
which result.
\end{abstract}
\end{center}
\vskip12pt

\setcounter{footnote}{0}
\renewcommand{\thefootnote}{\arabic{footnote}}
\vfill
\clearpage
\setcounter{page}{1}
\pagestyle{plain}

A complete theoretical description of $W$ and $Z$ boson
production at high-energy hadron colliders is necessary for
precision Standard Model phenomenology (for example, the measurement
of $M_W$) and for the reliable estimation of  backgrounds to  new physics
processes. An important issue is the calculation of the
transverse momentum ($q_T$) distribution at small $q_T$, which
requires  the resummation
of large `Sudakov' double logarithms.
These order $\alpha_S^n
\ln^{2n-1}(Q^2/q_T^2)$ contributions  arise from soft gluon emission
and lead to a breakdown of fixed-order perturbation theory
as $q_T \to 0$.

Although the resummation of the soft gluon contributions is achieved
most naturally in impact parameter (Fourier transform) space~\cite{CSS}, there
are certain advantages in performing the resummation directly in
transverse momentum space~\cite{EV}. For example, the matching with fixed-order
results at large $q_T$ is difficult in the impact parameter approach,
since this gives oscillatory behaviour for the large $q_T$ `tail'
of the transformed resummed soft gluon contribution.

The  resummation of the large logarithms directly in $q_T$ space
 has therefore received much
recent attention. In particular, two complementary
approaches have been  proposed in Refs.~\cite{FNR} (FNR) and~\cite{KS} (KS).
The aim of both these approaches is to improve the theoretical
description of the small  $q_T$ cross section by including certain
subleading (i.e. order $\alpha_S^n
\ln^{2n-r}(Q^2/q_T^2), \ r \geq 2$) logarithms in the resummation.
However the resulting predictions are qualitatively very different
at very  small $q_T$. The aim
of this note is to present and discuss the differences between
the FNR and KS
calculations.\footnote{We also briefly comment
on the related work of~\cite{EV} (EV).}
We pinpoint the exact set of subleading logarithms summed in the
two cases, and show how their inclusion
 can lead to significant differences in the predictions.

In order to highlight the differences  we will consider
both approaches in their most simplified version, i.e.
we shall restrict ourselves to the
parton-level subprocess cross section and ignore certain other subleading
corrections, as decribed below. We also do not consider  any non-perturbative
($q_T$ smearing) effects.

The springboard for both approaches is the general expression in impact
parameter ($b$) space for the vector boson transverse momentum
distribution in  the Drell-Yan process~\cite{CSS} at the quark level:
\be
{d\sigma \over d q_T^2} ={\sigma_0 \over 2}\int_{0}^{\infty} b d b  \, J_{0}( q_{\tiny T}
b) e^{S(b,Q^2)} \,,
\label{resum}
\ee
where
\ba
\label{eq:abseries}
S(b,Q^2) = - \int_{b_0^2 \over b^2}^{Q^2} \frac{d\bar\mu^2}{\bar\mu^2}
\bigg[ \ln \bigg ( {Q^2\over\bar \mu^2} \bigg ) A(\alpha_S(\bar\mu^2)) +
B(\alpha_S(\bar\mu^2)) \bigg ] \,,\label{Sbs} \\
A(\alpha_S) = \sum^\infty_{i=1} \left(\frac{\alpha_S}{2 \pi} \right)^i A^{(i)}\
, \quad
B(\alpha_S) = \sum^\infty_{i=1} \left(\frac{\alpha_S}{2 \pi} \right)^i B^{(i)}\
,\nn
\ea
\bann
b_0=2\exp(-\gamma_E)\,, \quad \sigma_0={4 \pi \alpha^2 \over 9 s} \,.
\eann
For the purposes of this simplified analysis we take the coupling $\alpha_S$ be
fixed
and retain only the leading
coefficient $A^{(1)}$, i.e.
\ba
A(\alpha_S) = \frac{\alpha_S C_F}{\pi} ,\ B(\alpha_S) = 0
\ea
with $C_F=4/3$. With these assumptions the $b-$space expression becomes
\be
{d\sigma \over d q_T^2} = {\sigma_0 \over 2}\int_{0}^{\infty} b d b \,  J_{0}( q_T  b) \exp
\bigg( -\frac{\alpha_s C_F}{2 \pi} \ln^2 \bigg ( \frac{Q^2 b^2}{b_0^2}
\bigg) \bigg) \ .
\label{eq:bspace}
\ee
We recall that the $b-$space formalism takes fully into account the
conservation of transverse momentum in multigluon emission,
$\delta(\vec{q_T} +\sum_i\vec{k_{Ti}} )$, and is therefore expected to provide
a better approximation of
the $q_T$ distribution at small $q_T$ than the Double Leading Logarithm
Approximation (DDLA), in which strong ordering of the gluons' $k_{Ti}$
is assumed. The DLLA leads to the Sudakov form factor expression~\cite{DDT}
\be
{1 \over \sigma_0}{d\sigma \over d q_T^2}\bigg|_{DLLA}= {d \over d q_T^2} \exp \left( -{\alpha_S C_F \over 2\pi}\ln^2
\left({Q^2 \over q_T^2} \right)\right) \ .
\ee
 The $b-$space
expression is (mathematically) well defined for all values of $q_T$.
In particular, in the
limit  $q_T \rightarrow 0$ it gives a finite  positive cross section, in
contrast to the DLLA result which, because of the vanishing of strong-ordered
phase space,  yields zero in this limit. However the large
$q_T$ behaviour of the $b-$space expression (particularly
when higher coefficients
$A^{(i)},\,B^{(i)}$ are taken into account) is not physical
-- the $q_T$ distribution oscillates between positive and negative values
due to the nature of the Bessel function.

After performing partial integration the expression (\ref{eq:bspace})
can be written (using the KS notation) as
\be
{1 \over \sigma_0}{d\sigma \over d \eta} = {d \over d \eta} \int_0^{\infty}
d\hat{b} J_1(\hat{b}) \exp \left( -{\lambda \over 2}
\ln^2 \left( {\hat{b}^2 \over \eta b_0^2} \right) \right)
\label{b_space}
\ee
with $\hat{b} = b q_T$, $\lambda = \frac{\alpha_S C_F}{\pi}$,
$\eta=\frac{q_T^2}{Q^2}$.

The next step in the KS approach  is to expand terms under the
integral in~(\ref{b_space}), which gives
\ba
{1 \over \sigma_0} {d \sigma \over d \eta} =
{1 \over \eta} \sum_{N=1}^{\infty} \lambda^N { (-1/2)^{N-1} \over (N-1)!}
\sum_{m=0}^{2N-1} 2^m \tau_m { \left( \begin{array}{c} 2N-1 \\ m \end{array} \right)}L^{2N-1-m} \,.
\label{qt_sum1}
\ea
Here $L=\ln\left({1 \over \eta}\right)=\ln\left({Q^2 \over q_T^2}\right)$
and the numbers $\tau_m$ are defined by~%
\footnote{The $\tau_m$ are called  $\bar{b}_m({\infty})$ in~\cite{KS}.}
\be
\tau_m \equiv \int_{0}^{\infty} d y J_{1}(y) \ln^m({y \over b_0})
\label{b_def}
\ee
and can be calculated explicitly using the generating function
\be
\sum_{m=0}^{\infty} {1 \over m!} t^m \tau_m =
e^{t \gamma_E}{\Gamma \left(1 + {t\over 2}\right) \over \Gamma \left(1 - {t\over 2}\right)} =
\exp \bigg[ -2 \sum_{k=1}^{\infty} { \zeta(2k+1) \over 2k+1} {\left( t \over
2 \right) }^{2k+1} \bigg] \,,
\label{b_form}
\ee
so that e.g. $\tau_0=1,\ \tau_1=\tau_2=0,\ \tau_3=-{1\over 2}\zeta(3)$
etc. For large $m$, the coefficients $\tau_m$ behave as $\tau_m \propto (-1)^m
m! 2^{-m}$, and the first twenty $\tau_m$ are tabulated in~\cite{KS}.

\begin{figure}[p]
\begin{center}
\hspace{-1cm}\mbox{\epsfig{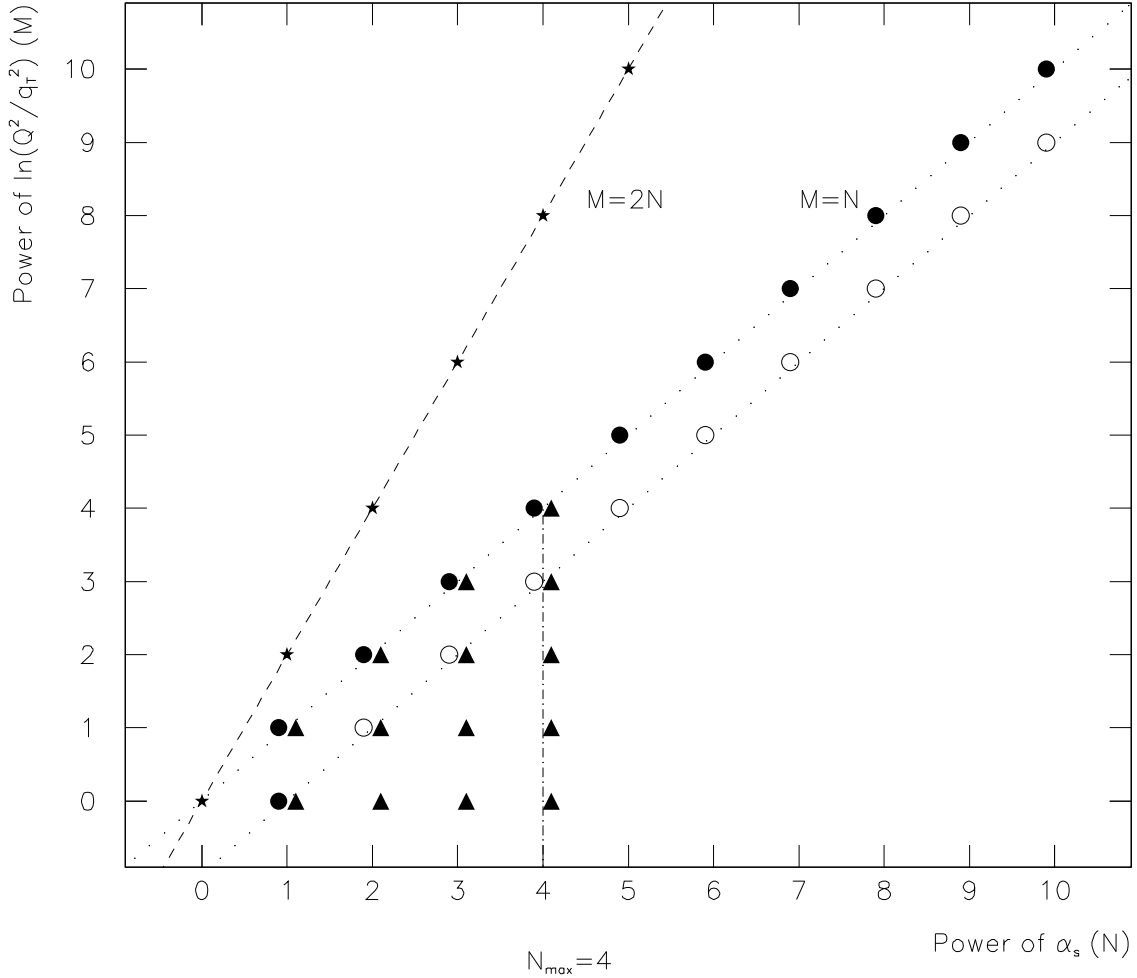}}
\end{center}
\caption{Schematic representation of contributions to~(\ref{FNR_sum})
  and to~(\ref{qt_sum2}). Circles correspond to the former expression,
  triangles to the latter one. An empty marker of a certain shape
  means that there exist other contributions in the perturbation series with the
  same power of $\alpha_S$ and $\ln(Q^2/q_T^2)$ which are not included in
  an expression coded with that shape.
  The points along the line labelled
                   `$M=2N$'
  represent terms coming from the Sudakov factor.}
\label{sum2_plot}
\end{figure}

By extracting the Sudakov form factor $\exp
\left(-{\lambda \over 2} L^2\right)$ the expression~(\ref{qt_sum1}) transforms into
\ba
{1 \over \sigma_0} {d \sigma \over d \eta} =
{\lambda \over \eta} e^{ -{\lambda \over 2} L^2}
\sum_{N=1}^{\infty} {(-2 \lambda)^{(N-1)} \over (N-1)!}
\sum_{m=0}^{N-1} { \left( \begin{array}{c} N-1 \\ m \end{array} \right)}
L^{N-1-m}
 \bigg[2\tau_{N+m}+ L \tau_{N+m-1}\bigg] \ .
\label{qt_sum2}
\ea
Naturally, for  numerical calculations based on the expression~(\ref{qt_sum2})
it is necessary to introduce a cut-off value $N_{\rm max}$. For example,
 Fig.~\ref{sum2_plot} shows the contributions to the double summation
 in (\ref{qt_sum2}) which are summed when
$N_{\rm max}=4$. Some illustrative numerical results based on the KS approach
will be presented below.

\begin{figure}[p]
\begin{center}
\hspace{-1cm}\mbox{\epsfig{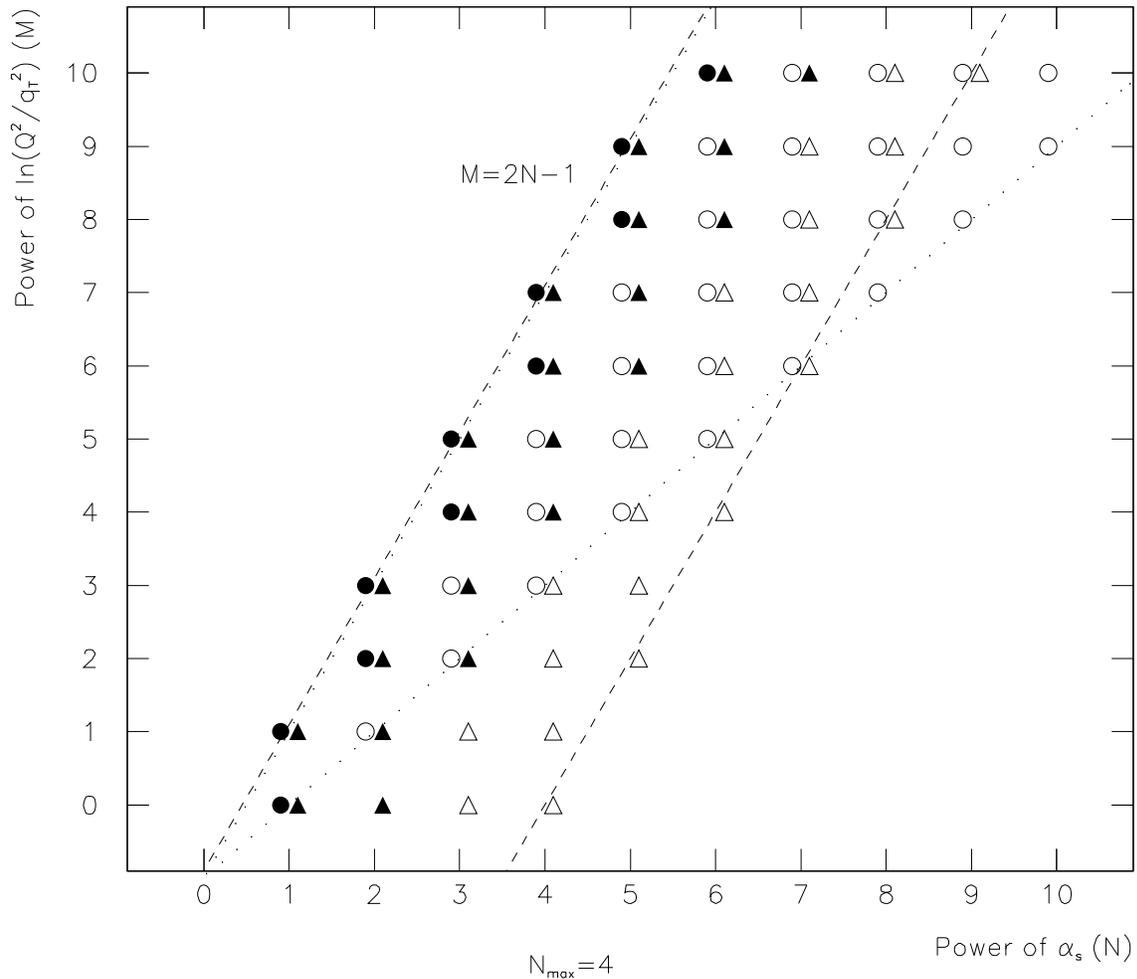}}
\end{center}
\caption{Schematic representation of contributions to~(\ref{FNR_sum}) with
  the Sudakov factor expanded and~(\ref{qt_sum1}). Circles correspond to
  the former expression, triangles to the latter one. An empty marker of a
  certain shape means that there exist other contributions in perturbation
  theory of the same power of $\alpha_S$ and $\ln(Q^2/q_T^2)$ which are not
  included in an approach coded with that shape.}
\label{sum1_plot}
\end{figure}

Another approach, suggested in~\cite{EV} (EV),
succeeds in developing an {\it analytic}
approximation when a slightly modified set of assumptions is considered. That
is to say, all the $\tau_m$ coefficients except  $\tau_0$ are set to be
zero and $B^{(2)}$ acquires an additional contribution
$-4 \tau_3 (A^{(1)})^2 =2\zeta(3)( A^{(1)})^2$. Including the expanded
 Sudakov factor, see Fig.~\ref{sum1_plot}, this corresponds to fully summing
 the first three leading series of logarithms, i.e. terms
of the form $\alpha_S^N L^{2N-1-m}$, $m=0,1,2$. In the EV approach,
the redefinition of $B^{(2)}$
correctly takes account of  the {\it first} term of the fourth series,  i.e. the
$\alpha_S^2 L^0$ term. On the other hand, it distorts other terms of this series
and terms from more subleading series wherever the  $B^{(2)}$
coefficient appears.

In the FNR approach~\cite{FNR}, one expands the exponent in~(\ref{b_space})
\be
 \exp \left( -{\lambda \over 2}
\ln^2 \left( {\hat{b}^2 \over \eta b_0^2} \right) \right)
= \exp \left( -{\lambda \over 2} \left( L^2
+\, 2 L\, L_b +\, L_b^2\right)\right)
\label{eq:expandexp}
\ee
where $L_b= \ln \left( {\hat{b}^2 \over  b_0^2} \right)$, and retains only the first two terms (`NLL approximation'):
\be
{1 \over \sigma_0} {d \sigma \over d \eta}\bigg |_{\cite{FNR}} =
{d \over d \eta} \int_0^{\infty}  d\hat{b} J_1(\hat{b})
\exp \left( -{\lambda \over 2}
L^2 - \lambda L\, L_b \right)
\label{FNR1}
\ee
Note that keeping only the leading $\sim L^2$ term in the exponent
corresponds  to the DLLA.
A great advantage of the `NLL approximation' is that the
  $\hat{b}$ integral can be performed
 analytically~%
\footnote{ Note that throughout this note the value of the lower limit
  of integration in~(\ref{Sbs}) is ${b_0^2 \over b^2}$. This is different
  from~\cite{FNR} where ${1 \over b^2}$ is chosen. Therefore the
  expression~(\ref{FNR2}) differs from the original expression in~\cite{FNR}
  by a constant.}:
\be
{1 \over \sigma_0} {d \sigma \over d \eta} \bigg |_{\cite{FNR}} =
{d \over d \eta} \left[
\exp \left( -{\lambda \over 2} L^2 \right)
\left( {2 \over b_0} \right)^{-2 \lambda L}
{\Gamma (1 - \lambda L) \over \Gamma (1 + \lambda L)}
\right]\,.
\label{FNR2}
\ee

There is in fact a direct link between the KS and FNR approaches.
If instead of
performing the integration in~(\ref{FNR1}) one expands the $b-$dependent terms
in the exponent and then performs the integration, the result is
\ba
{1 \over \sigma_0} {d \sigma \over d \eta}\bigg|_{\cite{FNR}} =
{\lambda \over \eta} e^{ -{\lambda \over 2}  L^2 }
\sum_{N=1}^{\infty} {(-2 \lambda)^{(N-1)} \over (N-1)!}
L^{N-1}  \left[ 2\tau_N + L \tau_{N-1}\right].
\label{FNR_sum}
\ea
Clearly this is just the expression~(\ref{qt_sum2}) taken at
$m=0$. Indeed the same result can be derived from the resummed
expression~(\ref{FNR2}) by recalling the definition of the generating
function~(\ref{b_form}) and using the relation
\be
\ln \Gamma (1+x) = -\ln(1+x) + x(1-\gamma_E) +\sum_{n=2}^{\infty} (-1)^n
[\zeta(n)-1]{x^n \over n}\,, \quad |x|<2 \,.
\ee
The contributions being resummed in both approaches are illustrated
schematically in Fig.~\ref{sum2_plot} and Fig.~\ref{sum1_plot}.
The aim of Fig.~\ref{sum2_plot} is to show which terms of the form
$\alpha_S^N L^M$ from the residual sums in~(\ref{qt_sum2})
and~(\ref{FNR_sum}) are taken into account. Here the FNR approach
corresponds to having two infinite lines of points (terms) while the KS approach
 results in a finite triangle of terms with size determined by $N_{\rm
  max}$. Terms emerging in the  {\it full} perturbative expansion, i.e.
  after expanding and multiplying in the
Sudakov factor, in both approaches are illustrated in Fig.~\ref{sum1_plot}.
Note that summing over all logarithmic terms with a given power of $\alpha_S$ must
result in the perturbative expansion coefficient of the same order, up to
logarithmic accuracy. Of course
a formula with an expanded Sudakov factor is valid only when
$\alpha_S L^2 \lapproxeq 1$.
The only reason for expanding the Sudakov factor here
is to determine which terms
in the overall perturbation series are actually being  resummed
in~(\ref{qt_sum2}) and~(\ref{FNR2}). It is these latter expressions, which can
be regarded as the `master equation' of the two approaches, that we use to
obtain numerical results, and both approaches remain
well-behaved provided $\alpha_S L \lapproxeq 1$.


\begin{figure}[p]
\begin{center}
\hspace{-1cm}\mbox{\epsfig{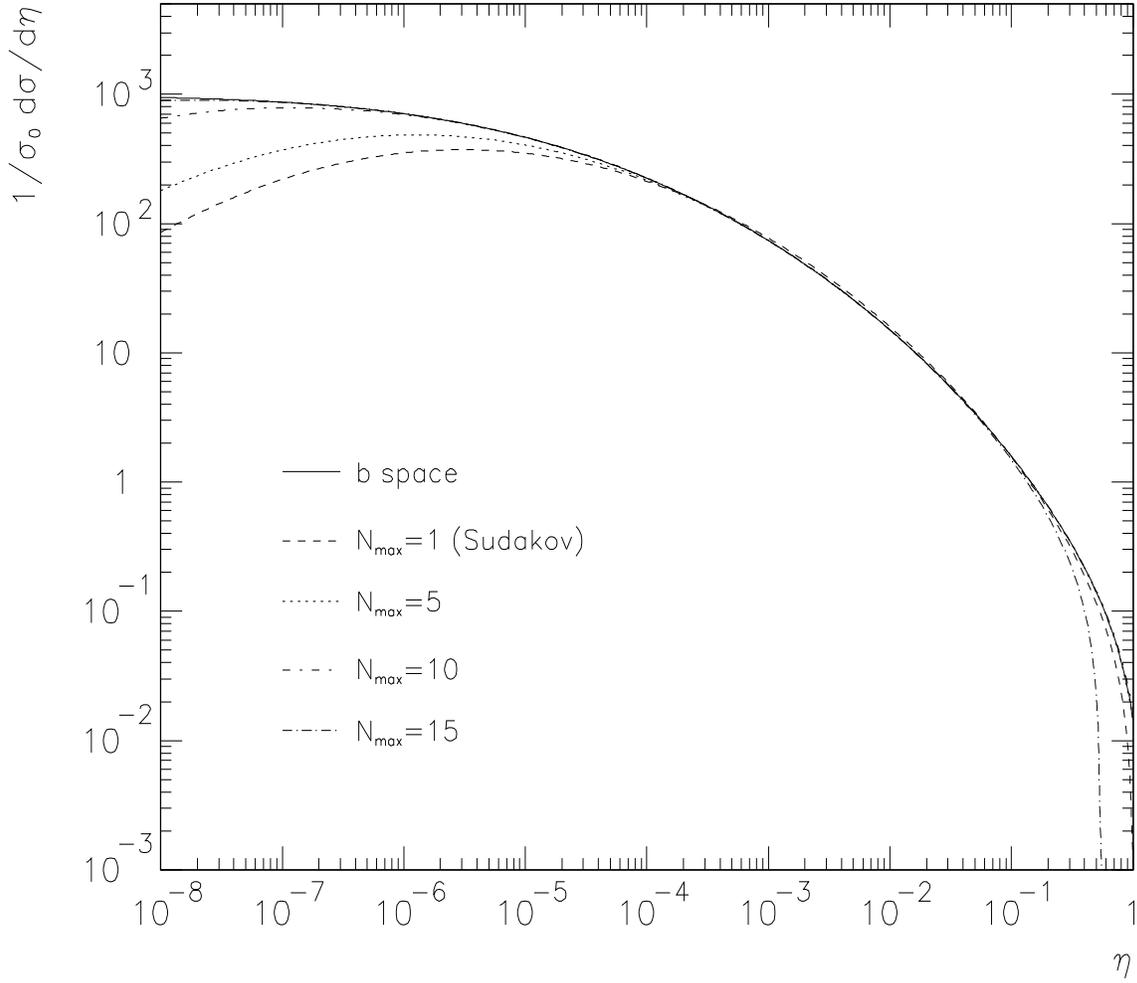}}
\end{center}
\caption{The $b-$space result compared to the expression~(\ref{qt_sum2}),
  calculated for various values of $N_{\rm max}$. Here $N_{\rm max}=1$
  corresponds to the DLLA approximation.}
\label{Nmax_KS}
\end{figure}

We next present some simple numerical comparisons
using both approaches~%
\footnote{The value of the strong coupling constant has
been fixed here at $\alpha_S$=0.2.}.
First we investigate the dependence of the KS result  on the point
of truncation $N_{\rm max}$. (Note that  $N_{\rm max}=1$ is
just the DLLA approximation.) It is clear from Fig.~\ref{Nmax_KS}
that for small values of $\eta$ (i.e. $q_T/Q \ll 1$)
the approximating of the $b-$space result
improves with  increasing $N_{\rm max}$. The `Sudakov dip'
at very small $\eta$ is more and more filled in as $N_{\rm max}$
increases, by contributions which are formally subleading in terms
of powers of $\alpha_S$ and $L$.

\begin{figure}[p]
\begin{center}
\hspace{-1cm}\mbox{\epsfig{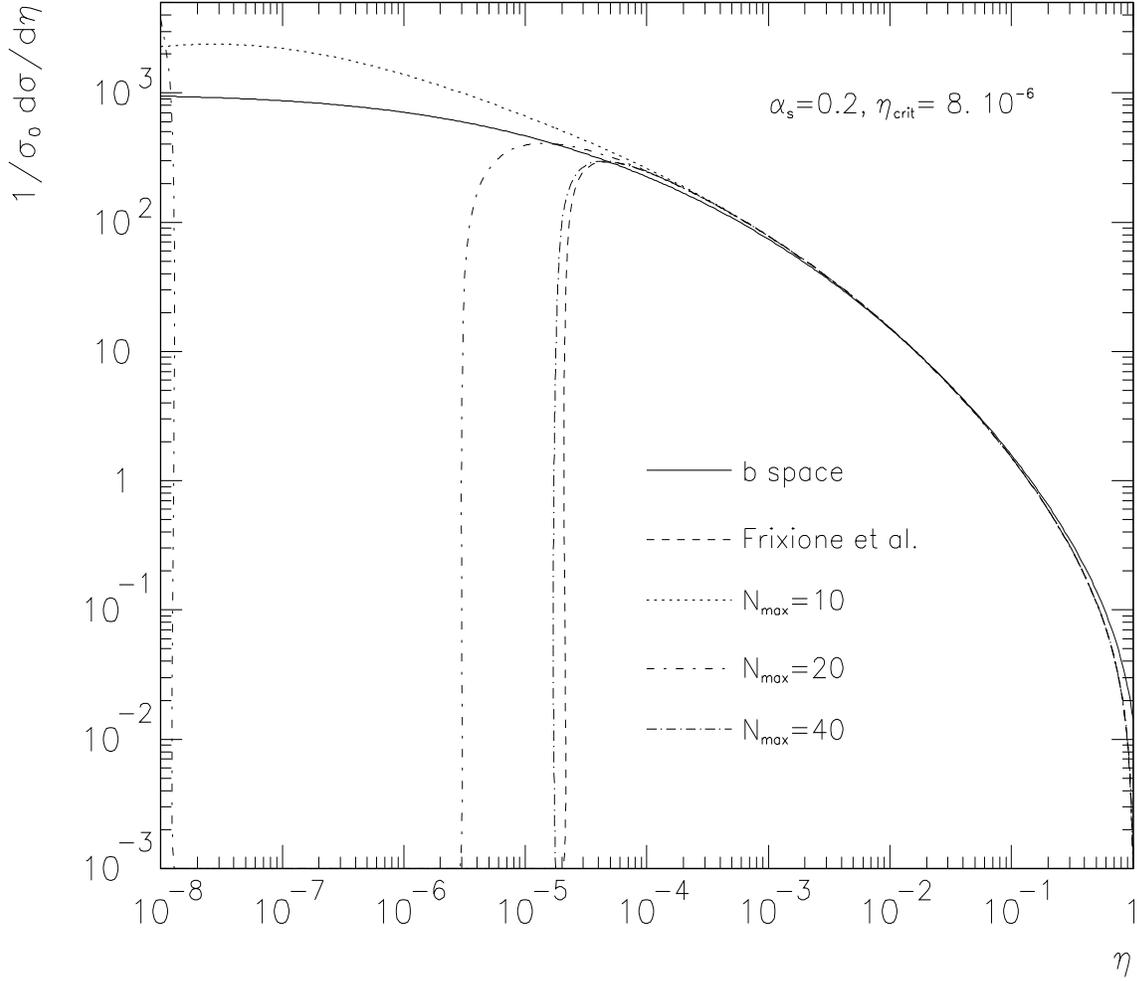}}
\end{center}
\caption{The $b-$space result compared to the
  expression~(\ref{FNR2})and~(\ref{FNR_sum}), calculated for various values of
  $N_{\rm max}$. With the choice $\alpha_s=0.2$,~(\ref{FNR2}) is only
  applicable for $\eta \gapproxeq 8. \times  10^{-6}$.}
\label{Nmax_FNR}
\end{figure}

On the other hand, the range of
applicability of the FNR resummed formula~(\ref{FNR2}) is seriously
restricted. As pointed out in~\cite{FNR}, the expression~(\ref{FNR2}) suffers
from singularities at $\lambda L= 1,2,...$. (In fact these
singularities are poles of order two.) Therefore the first pole encountered
as $\eta$ decreases is at $\eta^{\rm crit}=\exp\left(-{1\over\lambda}\right)$,
i.e. $q_T^{\rm crit}=Q\exp\left(-{\pi \over 2\alpha_S C_F}\right)$.~%
\footnote{This may appear to be an irrelevantly small value but,
as shown in~\cite{FNR}, when the running coupling constant is used
 the pole moves
  significantly towards higher values of $q_T$.}
Figure~\ref{Nmax_FNR} shows the resummed FNR
result~(\ref{FNR2}) as a function of $\eta$. The pole at  $\eta^{\rm crit}$
is evident (the distribution $\to -\infty$ as the singularity is approached
from above).  The resummed result is also compared to the `truncated'
expression~(\ref{FNR_sum}) for various values of the
 cut-off parameter $N_{\rm max }$. This shows the effect of successively
 adding more and more of the subleading `$m=0$' terms along the
 two infinite lines of Fig.~\ref{sum2_plot}, starting from the Sudakov expression
 for $N_{\rm max } = 1$.   Convergence to the (singular) resummed FNR
 result ~(\ref{FNR2})  for large  $N_{\rm max }$ is clearly evident.
\begin{figure}[p]
\begin{center}
\hspace{-1cm}\mbox{\epsfig{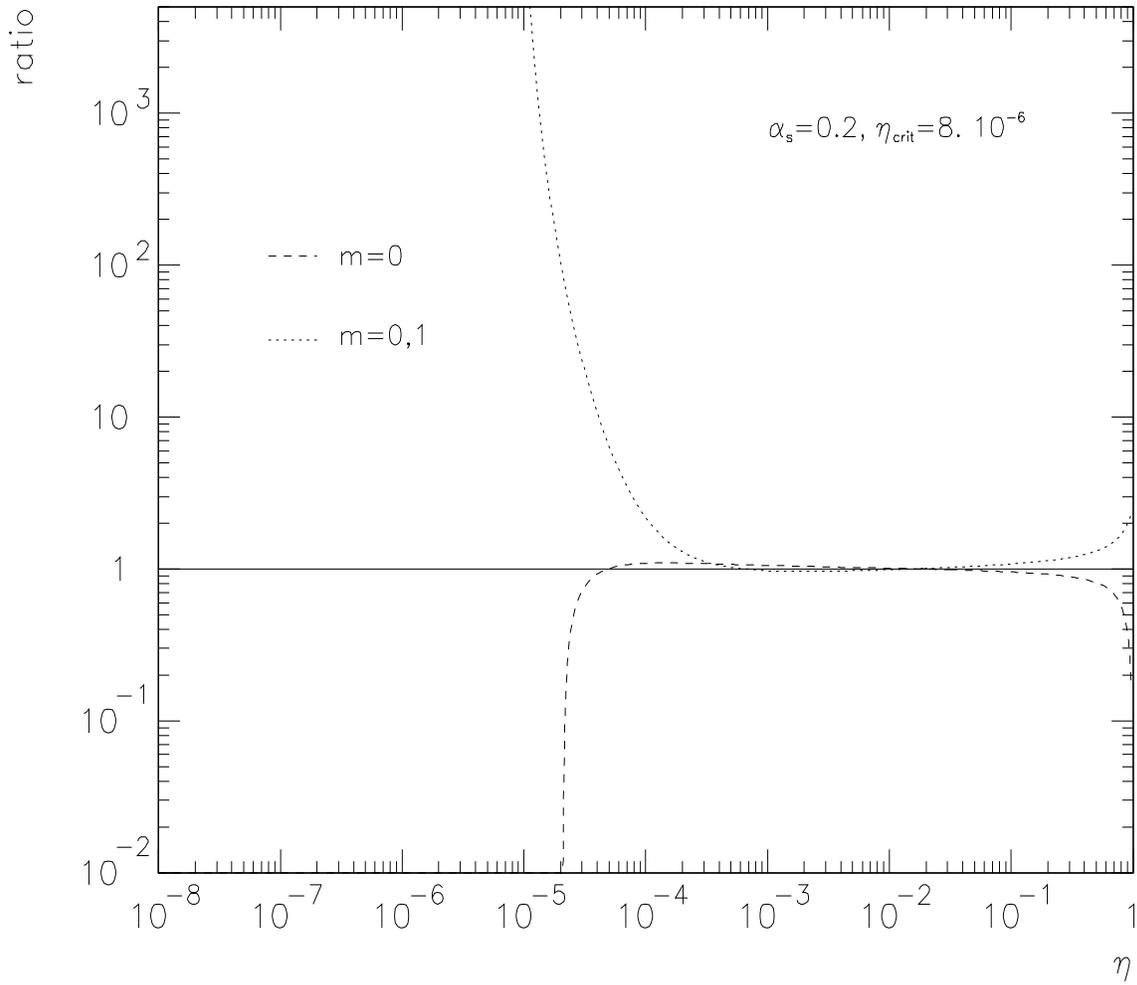}}
\end{center}
\caption{The ratio of the numerically calculated~(\ref{FNR2}) (m=0 curve)
  and~(\ref{m01}) (m=0,1 curve) to the $b-$space result.}
\label{m_ratio}
\end{figure}

A natural extension of the approach of~\cite{FNR} would be a resummed
analytic expression also including $m=1$ terms in the classification
of~(\ref{FNR_sum}). In fact one can systematically include the subleading
`NNLL' terms of (\ref{eq:expandexp}) using the identity
\be
 \exp  \left( -{\lambda \over 2} \left( 2 L\,  L_b
\, +\,  L_b^2
 \right)\right)
=
\sum_{j=0}^{\infty} \, {1 \over j! } \, \left( -{\lambda \over 2} \right)^j
\, { d^{2j} \over d (\lambda L)^{2j} } \,
\exp  \left( -\lambda  L \, L_b
 \right)
\ee
which generates more subleading terms as derivatives of the FNR analytic `NLL' result.
In particular, including the $m=0,1$ contributions yields
\ba
{1 \over \sigma_0}{d \sigma \over d \eta} \bigg |_{m=0,1} =
{d \over d \eta} \left\{
\exp \left( -{\lambda \over 2} L^2 \right)
\left(1- {\eta^2 \over 2 \lambda}{d^2 \over d^2 \eta} - {\eta \over 2 \lambda}  {d \over d \eta}\right)
\left[\left( {2 \over b_0} \right)^{-2 \lambda L}
{\Gamma (1 - \lambda L) \over \Gamma (1 +\lambda L)}
\right] \right\} .
\label{m01}
\ea
This, however, does not cure the singularity problem but
makes it even worse. It turns out that if the upper limit of the sum over $m$
increases by 1, the order of the poles increases by 2, e.g. the
formula~(\ref{m01}) has poles of order four at $\eta^{\rm crit}$.
Moreover, as this upper limit
increases, the region where the approximation
 of the $b-$space result
becomes better, contracts. This is illustrated in
Fig.~\ref{m_ratio}, where we show the
ratio of the numerically calculated expression~(\ref{FNR2}),~(\ref{m01})
and the `full' $b-$space result.

The authors of~\cite{FNR} argue that the subleading terms in the
original expression~(\ref{b_space}) possess a divergent behaviour. These
terms have been shown to have factorially growing coefficients and originate
from the small-$\hat{b}$ region of integration. Their presence manifests
itself in the  KS approach  which takes some of those subleading
terms into account. Indeed for large $N_{\rm max}$ (i.e. more subleading
terms), the presence of factorially growing subleading
coefficients  can be seen at large $\eta$,
e.g. in  Fig.~\ref{Nmax_KS}, where the cross section
can become negative.  But this behaviour is not entirely
unexpected. Let us recall  that the $b-$space
formalism was invented to provide a good description in the {\it small} $\eta$
regime where $\alpha_S \ln(1/\eta)$ becomes large. Thus the presence of
factorially growing terms which manifest themselves for large $\eta$ can be
understood as an artefact of the $b-$space method. In fact, in this
formalism, the recovery of a credible theoretical result in the large $\eta$
domain relies on  careful matching with fixed-order perturbation theory.

In summary, we have shown that the KS and NFR approaches start from the
same expression for the cross section in $b$ space, but organise the
perturbative expansion in $q_T$ space in different ways such that different
subleading  $q_T$ logarithms are included.  In the KS approach, `towers' of
subleading logarithms fill in the Sudakov (DLLA) dip at small $q_T$. In the FNR
approach, a particular `subset' of subleading logarithms is resummed
to all orders, but the resulting expression  has a singularity at
$q_T = q_T^{\rm crit}$, below which the cross section is not defined. The FNR
result can be obtained in the KS approach by including only the appropriate
subleading terms.

The basic question remains as to whether
 there is a definite value of $q_T$ below which the perturbative
  expression for the cross section cannot be calculated.
The original argument
for the $b$-space approach was that it allowed a non-zero cross section
at $q_T = 0$ to be generated by the emission of soft gluons whose transverse
momentum vectors cancelled, a phase space region clearly outside the
strongly-ordered DLLA domain. The KS approach is designed to take these
(formally subleading) contributions into account in a systematic way.
The `price' one pays is a series with factorially growing coefficients
that drive the behaviour at large $q_T$, but this is in any case
outside the region of applicability
of the whole approach.
In contrast, the validity of the
FNR approach seems to depend on the extent to which the `LL' and `NLL'
terms as defined in Eq.~(\ref{eq:expandexp}) do
actually give the dominant contribution to the small $q_T$ cross section.
Since we have shown that attempts to systematically
 include the `NNLL' contributions in this approach
 lead to even more singular behaviour than the one observed in the `NLL' case,
this may cast doubt on the validity of the NLL
 approximation. In any case, one incontrovertible conclusion is that
 this is an interesting and important issue that deserves further study.

\vskip 0.5truecm

\noindent{\bf Acknowledgements} \\
We thank Stefano Frixione, Paolo Nason and Giovanni Ridolfi
for useful and informative discussions.
This work was supported in part by the EU Fourth Framework Programme `Training and Mobility of Researchers', Network `Quantum Chromodynamics
and the Deep Structure of Elementary Particles', contract FMRX-CT98-0194 (DG 12 - MIHT).
A.K. gratefully acknowledges financial support received from the ORS Award
Scheme and the University of Durham.

\pagebreak

\end{document}